\documentclass[manuscript]{aastex}

\slugcomment{To be submitted for publication}
\shorttitle{Time series analysis of gamma-ray blazars}
\shortauthors{Nakagawa and Mori}

\begin{document}

\title{Time series analysis of gamma-ray blazars and implications for the central black-hole mass}
\author{Kenji Nakagawa and Masaki Mori}
\affil{Department of Physical Sciences, Ritsumeikan University, Kusatsu, Shiga 525-8577, Japan}

\begin{abstract}

Radiation from the blazar class of active galactic nuclei (AGN) 
exhibits fast time variability which is usually ascribed
to instabilities in the emission region near 
the central supermassive black hole.
The variability time scale is generally faster in higher energy region, 
and data recently provided by the {\it Fermi} Gamma-ray Space Telescope 
in the GeV energy band enable a detailed study of the temporal behavior of AGN.
Due to its wide field-of-view in the scanning mode, most 
sky regions are observed for several hours per day and daily 
light curves of many AGN have been accumulated for more than 4 r.

In this paper we investigate the time variability of 
15 well-detected AGNs by studying the normalized power spectrum density 
of their light curves in the GeV energy band. 
One source, 3C 454.3, shows a specific time scale of $6.8\times10^5$~s, 
and this value suggests, assuming the internal shock model,  
a mass for the central black hole of $(10^8$--$10^{10})M_\odot$ 
which is consistent with other estimates.
It also indicates the typical time interval of ejected blobs 
is $(7$--$70)$ times the light crossing time of the Schwarzschild radius.

\end{abstract}


\section{Introduction}\label{sec:Intro}

About 1\,\% of galaxies have an active galactic nucleus (AGN),
which emits $10^9$--$10^{14}$ times the solar power over a wide range 
of the energy spectrum,
from radio to gamma-ray energies.
AGNs constitute one of the most violently variable and interesting
classes of object in the universe.
The activity of AGN is believed to originate from the central supermassive black hole, with
masses of $10^6$--$10^9$ solar mass ($M_\odot$), 
and part of their energy is emitted in electromagnetic radiation
from the surrounding region including the accretion disk formed around the black hole 
and relativistic jets ejected along rotation axes \citep[e.g.,][]{Urry1995}.
A subclass of radio-loud AGN are called blazars, in which the line of sight lies
close to the jet axis, and the emission from relativistic jets is only visible
in this class of AGN due to the relativistic beaming effect, especially in the
high-energy region.
The electromagnetic spectra of blazars are dominated by non-thermal radiation 
produced in the jets.
The popular scenario to explain these emission spectra assumes that
the particles in the jets are accelerated to high energies by diffusive 
shocks in the jets and induce emission via interaction with 
surrounding matter/radiation \citep[e.g.,][]{Fossati1998}%
\footnote{However, there are
challenges to this standard view based on recent observations; for examples,
the gamma-ray flares from 3C 279 showing strong optical polarization 
\citep{Abdo2010a}, and
the extremely rapid dissipation, of the order of few hours or minutes 
\citep[e.g.,][]{Aharonian2007,Aleksic2011}.}.

Observations of blazars at various wavelengths have revealed fast time variability
which is most plausibly related to instabilities in the emission environment near the
black hole \citep[e.g.,][]{Ulrich1997}.
Past observations suggest the variability is larger at higher energies
\citep[the most extreme example is Mrk 501][]{Nowak2012},
which may indicate the higher energy emission comes from the region
closer to the central black holes.
The variability time scale reflects the size of the emission region,
and thus the study of temporal behavior of gamma-ray flux is an excellent
probe of the region close to the central engine, i.e., the
supermassive black hole.

Blazars are known to show flaring activity which occurs randomly and continues for
several days to months. In order to study their temporal variability precisely, 
blazars should be monitored continuously, or at least frequently. 
It is not an easy task for narrow field-of-view telescopes like optical, X-ray,
and Cherenkov (TeV gamma-ray) instruments to monitor many blazars for long periods.
Besides, their observations are limited as ground-based
telescopes can only operate on dark, clear nights and X-ray satellites are used in
pointing-mode observations.

Nevertheless, in the X-ray band, \citet{Hayashida1998} 
evaluated the central black hole masses in several AGNs based on their 
rather well-sampled X-ray light curves obtained with the {\sl Ginga} satellite
and suggested the masses 1$\sim$2 orders of magnitude smaller than previous estimates.
\citet{Kataoka2001} studied time variability of
three TeV-detected AGNs based on {\it ASCA} and/or {\it RXTE} observations
and showed that $(10^{7}\sim 10^{10})M_\odot$ black holes and internal shocks
that start to develop at 100 times the Schwarzschild radii could explain the observed
properties.

In the GeV gamma-ray band, the {\sl Fermi} Gamma-ray Space Telescope has
been monitoring the whole sky with the Large Area Telescope (LAT) since 2008.
The LAT is a wide field-of-view gamma-ray imager that observes one-fifth of the sky
at any instant and that scans the whole sky in a day 
\citep{Atwood2009}.
The second {\sl Fermi}-LAT catalog,
which contains 1092 (28 identified and 1064 associated) AGN
among the 1873 detected sources in the 100\,MeV to 100\,GeV range \citep{Nolan2012}.
Gamma-ray light curves of several tens of blazars are provided on a daily 
basis and this is a good database to study the time variability of blazars.

\citet{Abdo2010b} reported a detailed analysis of the variability of 106 objects 
in the {\sl Fermi}-LAT Bright AGN Sample. They showed that the temporal
behavior of gamma-ray fluxes of variable sources
can be described by power-law power spectral density (PSD) in general, 
with a few blazars that showed strong activity exhibiting complex and
structured temporal profiles. They examined whether it was possible to characterize
blazar type with the PSD slope, but the results were not conclusive.

In this paper, we report the time series analysis of gamma-ray light curves
of 15 blazars based on {\sl Fermi}-LAT data and discuss the results in relation to
the properties of the central engine.
Our analysis is the first systematic study of long-term variability of
blazar emission in the gamma-ray energy band, although the 
analysis method itself has been applied and reported previously
\citep[(e.g., in the X-ray band,][]{Lawrence1987, McHardy1987, Miyamoto1994, Hayashida1998, Kataoka2001}.

\section{Data and Analysis}

We use the ``monitored source light curves'' provided by 
the Fermi Science Support Center (FSSC)%
\footnote{\url{http://fermi.gsfc.nasa.gov/ssc/data/access/lat/msl\_lc/}\,\,
 Note that, as stated here, these light curves are
 preliminary and fluxes do not have
 absolute calibration, and a preliminary instrument response function
 is used. }
for bright and transient gamma-ray sources.
They are regularly updated throughout the mission.
In this paper, we analyze the daily light curves of 15 AGN
(Table \ref{tab:PSD_result}) in the energy range
 100~MeV -- 300~GeV. The data period is between 2008 August 9
and 2012 April 26.
These sources are selected because a large fraction of data points are detections,
not upper limits, so that we can extract useful information on time variability.
Note that the usage of daily light curves spanning 44 months 
naturally limits our time series analysis to the $10^{-8}$--$10^{-5}$\,Hz range. 
Faster variability observed in the case of 3C 454.3 \citep{Abdo2011}, 
for example, is out of the scope of the present analysis.

As we noted in Section \ref{sec:Intro}, the variability time scale of AGN
flares reflects the size of the emission region,
and thus the study of temporal behavior of gamma-ray flux can be a good
probe to explore the physical environment close to the central engine.
However, characterizing the time scale is not a simple task
since we know that the intensity of gamma-ray emission from AGN flares 
varies very irregularly.
The fastest doubling time has been widely used \citep[see][for example]{Barr1986}
as a variability measure, but it depends on data quality and coverage.
Here we adopt a spectral analysis, the normalized power spectrum density (NPSD),
to evaluate the characteristic timescale of light curves 
which fluctuate chaotically, after \citet{Miyamoto1994}.

\subsection{Normalized Power Spectral Density} \label{sec:PSD}

The PSD shows the degree of variation at every frequency 
(or cycle) by calculating the Fourier transform of time variable data 
\citep{Lawrence1987, McHardy1987}.

The NPSD, which is obtained by dividing the
PSD by the average source intensity squared, has proven to be useful
to compare variability at each frequency even if the brightness changes
\citep{Miyamoto1994, Kataoka2001}. 
It is defined as
 \begin{eqnarray}
    P(f) &=& \frac{[a^2(f)+b^2(f)-\sigma^2_{\mathrm{stat}}/n] T}{F^2_{\mathrm{av}}} 
      \label{eq:npsd}\\
    a(f) &=& \frac{1}{n} \sum^{n-1}_{j=0}F_j \cos(2 \pi f t_j) \\
    b(f) &=& \frac{1}{n} \sum^{n-1}_{j=0}F_j \sin(2 \pi f t_j) 
 \end{eqnarray}
where $F_j$ is the source count rate at time $t_j$ ($0 < t_j< n-1$), 
$T$ is the total time length, 
$F_{\mathrm{av}}$ is the mean value of source count rates, 
and $\sigma_{\mathrm{stat}}$ is the error due to counting statistics.
In our analysis, we calculated the power $P(f)$ for some discrete
frequencies given by $f=k/T$ ($k$ is an integer and $1 < k < n/2$) and averaged.
The {\sl Fermi}-LAT light curves are given with flux errors ($e_j$)
and their standard deviation ($\sqrt{\sum_j e_j^2/N}$) 
is substituted for $\sigma_{\mathrm{stat}}$
as a rough estimate of error of counting statistics (see Section \ref{sec:PN} 
for more discussion).
The error bars of the NPSD are standard deviations of powers in each
frequency bin
\citep[see][for more discussion of NPSD analyses]{Hayashida1998,Kataoka2001}. 
In our calculation of the NPSD, we did not use upper
limits in the light curves.
In addition, we did not interpolate any blank (i.e., missing) data contained in the light curves.

\subsection{Poisson Noise} \label{sec:PN}

If the time series is a continuous counting rate binned into
intervals, as it is here, the effect of Poisson noise
is to add an approximately constant amount of power
to the NPSD at all frequencies. At high frequencies,
where the counting rate is low, the NPSD will be
dominated by the flat (white) Poisson noise spectrum.
In the definition of NPSD (Equation \ref{eq:npsd}) this noise
is subtracted in the term $\sigma^2_{\mathrm{stat}}/n$ \citep[see][for further discussion]{Vaughan2003}.

In the present case, $F_j$ ($e_j$)
is calculated as the counts (count error) divided by the exposure,
and the daily exposure is almost uniform for the {\sl Fermi}-LAT
observations. The Poisson noise is therefore approximately subtracted
in the calculation of the NPSD (Equation \ref{eq:npsd}).
This treatment formally assumes zero background flux, but
this is a reasonable approximation in our case, since the
{\sl Fermi}-LAT light curves are released after 
subtracting background counts when they are processed at the FSSC.
If the subtraction of the Poisson noise estimated by the quoted 
flux errors is not sufficiently accurate, there will be a
residual constant which becomes dominant at high frequencies in the NPSD,
which can be seen in some cases in our results (next section).

\section{Results} \label{sec:results}

We calculated the NPSD for the {\sl Fermi}-LAT daily light curves of 15 AGNs
using 9 frequency bins divided logarithmically from $10^{-7}$~Hz
to $10^{-5.2}$~Hz. Plots are shown in Figure~\ref{fig:PSD}.

One may note that data points at high frequencies are missing
for most of the sources in Figure~\ref{fig:PSD}. 
One reason is that many observations in the {\sl Fermi}-LAT light
curves yield only upper limits.
Another reason is that large flares which last for ten of days or more are rare:
we cannot have points above $\sim 10^{-6}$~Hz without such flares.

We applied least-square fits to these points assuming a power-law 
 \begin{equation}
  f(\nu) \propto \nu^{\gamma}
 \end{equation}
and/or a broken power-law
 \begin{equation}
  f(\nu) \propto
   \cases{
     \nu^{\gamma_1} \ (\nu < \nu_{b}) \cr
     \nu^{\gamma_2} \ (\nu \geq  \nu_{b}), \cr
   }
 \end{equation}
where $\nu$ is the frequency and $\nu_b$ is the ``turnover'' frequency.
Fit parameters are summarized in Table~\ref{tab:PSD_result}.
The fit lines are overplotted in Figure~\ref{fig:PSD}, where
broken power-law lines are plotted only when the reduced $\chi^2$ values
are smaller than single power-law values.. 
We see NPSDs for four sources, PKS 0537$-$441, 3C\,279,
3C\,454.3 and PKS 2326$-$502, are better fitted by broken
power-laws than by single power-laws.

The NPSD plots for PKS 0537$-$502, 3C 279 and PKS 2326$-$502 show upward turnovers 
above $10^{-6.18}$~Hz, $10^{-5.88}$~Hz and  $10^{-6.10}$~Hz, respectively,
but the slopes above these frequencies ($\gamma_2$) have large uncertainties
and are consistent with zero:
thus they may have reached a constant Poisson-like noise level
which is not removed by our rough estimate of counting statistical error
(see section \ref{sec:PSD}).
We checked the difference of NPSD values 
before and after
removing the Poisson noise, assuming
the value of Poisson noise is the normalized square-root of
sum of squares of several flux's error: $\sqrt{\sum e^2_i/N}$.
With this procedure,
the NPSD values at high frequencies showed smaller values than those before removal,
but the slope above the turnover of the NPSD plot remained flat
(consistent with zero slope).
Thus, even if we could not remove the effect of Poisson noise completely,
it seems this flattening behavior does not have a physical origin. 

On the other hand, the NPSD plot for 3C 454.3 show a turnover at $10^{-5.83}$~Hz
and the slope above it, $-3.08\pm 0.83$, is well determined.
The reduced $\chi^2$ value decreases from 1.29 for single power-law
fit, which is not at acceptable level, to 0.23 for broken power-law fit, which
is acceptable.
Thus, only the plot for 3C\,454.3, which exhibited an extraordinary large flare in
2010 November \citep{Abdo2011}, seems to show a physically meaningful 
turnover, at $10^{-5.83}$~Hz, which we discuss further in the next section.

\begin{deluxetable}{cccccccccccc}
\tabletypesize{\scriptsize}
\rotate
\tablecaption{List of Aanalyzed AGNs and Results of PSD Fitting. $N$ is the number of 
observations (upper limits are not included). See text for details.}
\tablewidth{0pt}
\tablehead{
\colhead{Source} & \colhead{$\alpha$ (2000)} & \colhead{$\delta$ (2000)}
 & \colhead{Redshift} & \colhead{$N$}
 & \multicolumn{2}{c}{Power-law} &{\,}& \multicolumn{4}{c}{Broken Power-law} \\
 \cline{6-7} \cline{9-12}
 \colhead{Name} & \colhead{(deg)} & \colhead{(deg)} & & 
 & \colhead{$\gamma$} & \colhead{$\chi^2/$d.o.f} &{\,}& \colhead{$\gamma_1$}
 & \colhead{$\gamma_2$} & \colhead{$\log_{10}\nu_b$} & \colhead{$\chi^2/$d.o.f}
}
\startdata
3C 66A              & 36.665 &   43.035 &0.444 &174&$-0.604\pm0.438$&0.121&&                &                &               \\
4C +28.07           & 39.468 &   28.802 &1.213 &104&$ 0.928\pm0.228$&0.464&&                &                &               \\ 
PKS 0426$-$380      & 67.168 &$-$37.939 &1.110 &220&$-1.160\pm0.470$&0.721&&                &                &               \\
PKS 0454$-$234      & 74.263 &$-$23.414 &1.003 &244&$-0.777\pm0.274$&1.268&&                &                &               \\
PKS 0537$-$441      & 84.710 &$-$44.086 &0.894 &509&$-0.555\pm0.239$&0.128&&$-0.863\pm0.631$&$-0.082\pm0.944$&$-$6.183&0.072 \\
S4 1030+61          &158.464 &   60.852 &1.401 & 73&$-0.012\pm0.454$&0.797&&                &                &               \\
Mrk 421             &166.114 &   38.209 &0.031 &498&$-0.384\pm0.205$&0.131&&                &                &               \\ 
PKS 1222+216        &186.227 &   21.380 &0.432 &348&$-0.648\pm0.212$&0.538&&                &                &               \\
3C 273              &187.278 &    2.502 &0.158 &251&$-1.301\pm0.265$&0.171&&                &                &               \\
3C 279              &194.047 & $-$5.789 &0.538 &367&$-1.078\pm0.246$&0.130&&$-1.231\pm0.352$&$ 0.976\pm5.496$&$-$5.879&0.125 \\
PKS 1510$-$089      &228.211 & $-9$.100 &0.360 &572&$-1.101\pm0.298$&0.249&&                &                &               \\
PKS 2155$-$304      &329.717 &$-$30.226 &0.116 &319&$-0.577\pm0.332$&0.308&&                &                &               \\
BL Lac              &330.680 &   42.278 &0.069 &143&$-0.412\pm0.469$&0.218&&                &                &               \\
3C 454.3            &343.491 &   16.148 &0.859 &837&$-1.498\pm0.164$&1.292&&$-0.999\pm0.235$&$-3.079\pm0.826$&$-$5.834&0.228 \\
PKS 2326$-$502      &352.337 &$-$49.928 &0.518 &247&$-0.922\pm0.204$&0.601&&$-1.257\pm0.436$&$ 0.089\pm0.886$&$-$6.103&0.487 \\
\enddata
\label{tab:PSD_result}
\end{deluxetable}

\begin{figure}
 \begin{tabular}{cc}
  \begin{minipage}{0.5\hsize}
   \begin{center}
    \includegraphics[width=6cm]{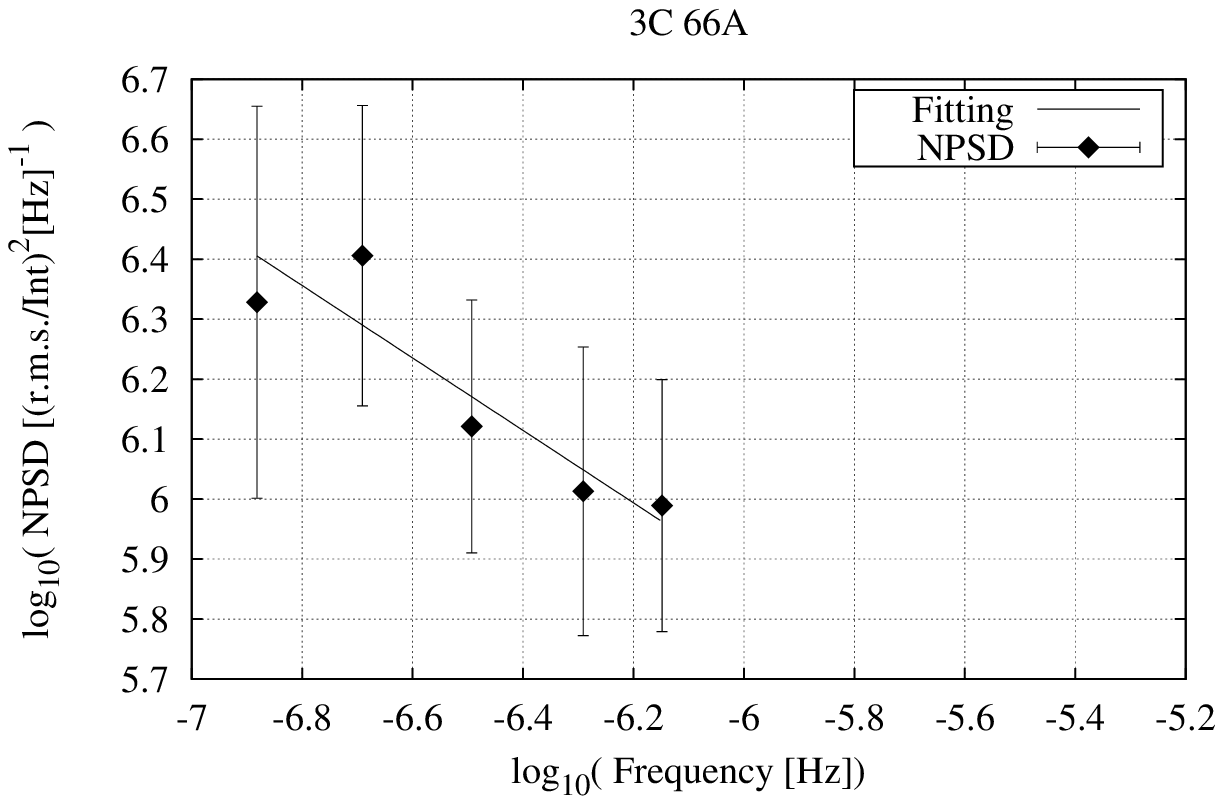}
   \end{center}
  \end{minipage}
  \begin{minipage}{0.5\hsize}
   \begin{center}
    \includegraphics[width=6cm]{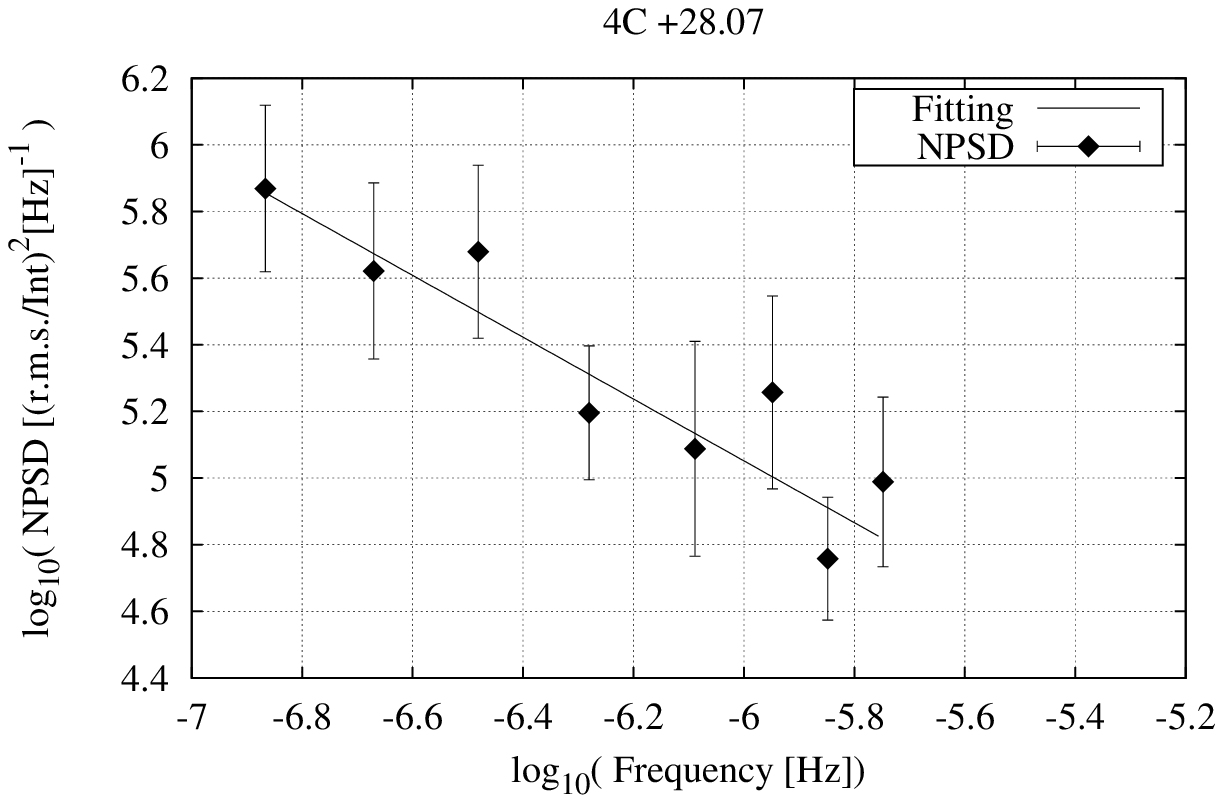}
   \end{center}
  \end{minipage}
 \end{tabular}
 \begin{tabular}{cc}
  \begin{minipage}{0.5\hsize}
   \begin{center}
    \includegraphics[width=6cm]{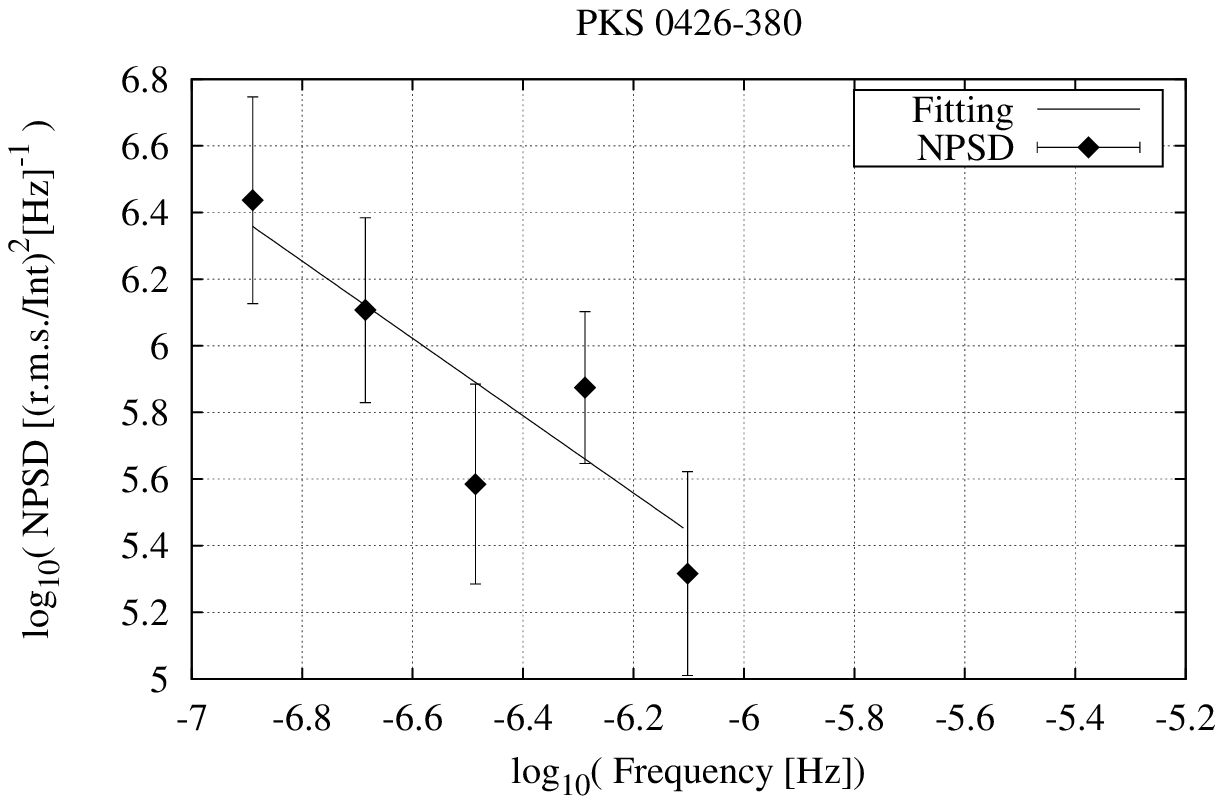}
   \end{center}
  \end{minipage}
  \begin{minipage}{0.5\hsize}
   \begin{center}
    \includegraphics[width=6cm]{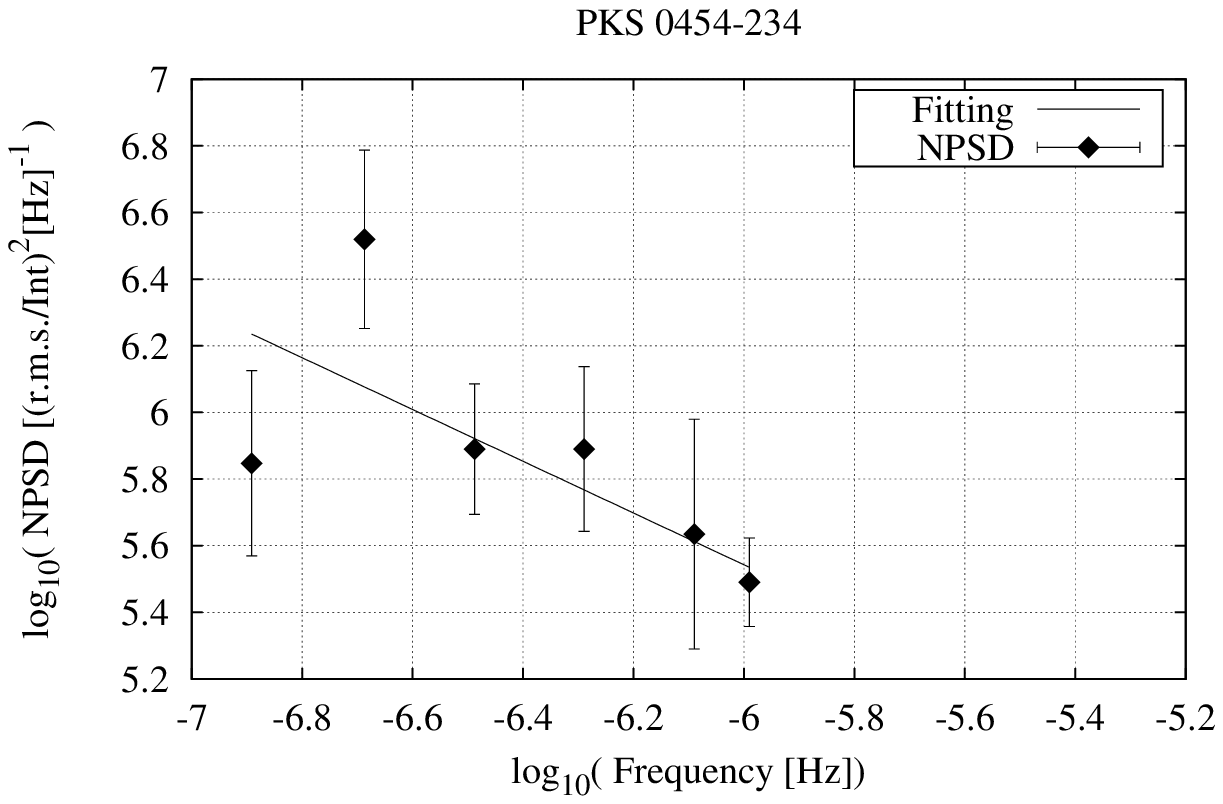}
   \end{center}
  \end{minipage}
 \end{tabular}
 \begin{tabular}{cc}
  \begin{minipage}{0.5\hsize}
   \begin{center}
    \includegraphics[width=6cm]{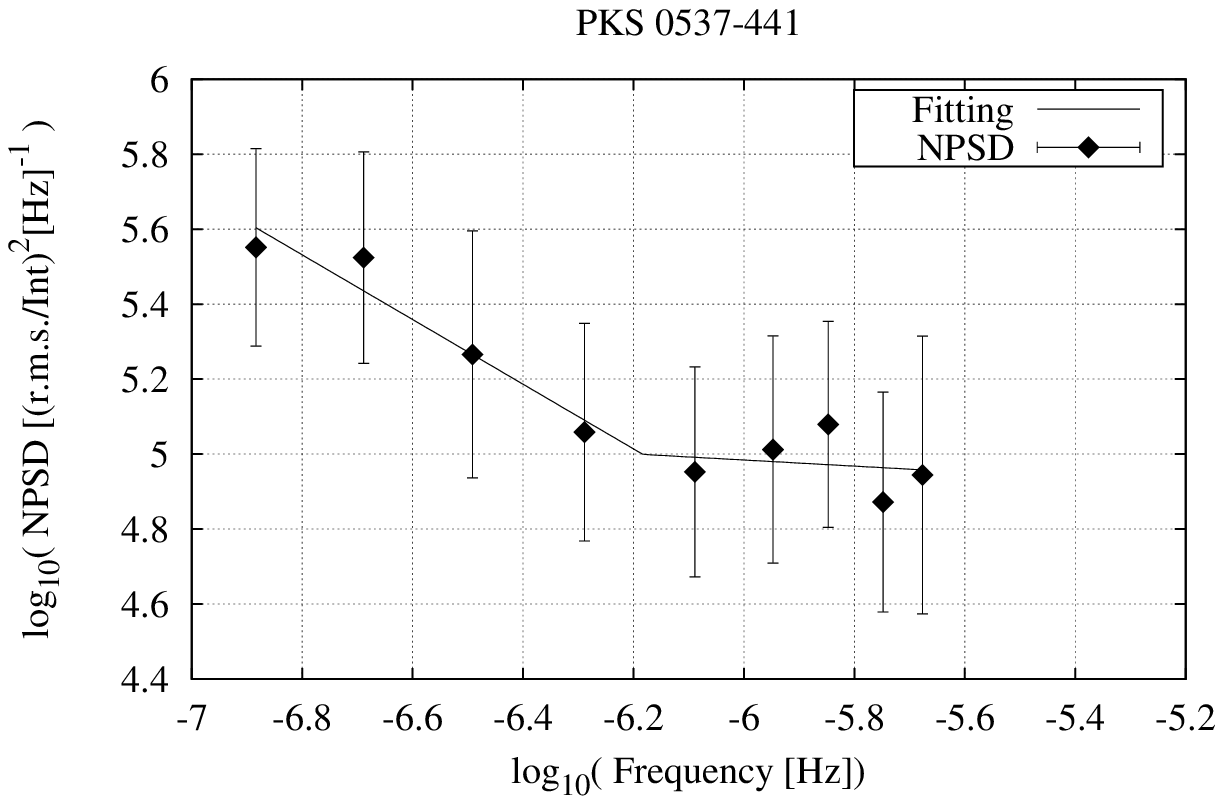}
   \end{center}
  \end{minipage}
  \begin{minipage}{0.5\hsize}
   \begin{center}
    \includegraphics[width=6cm]{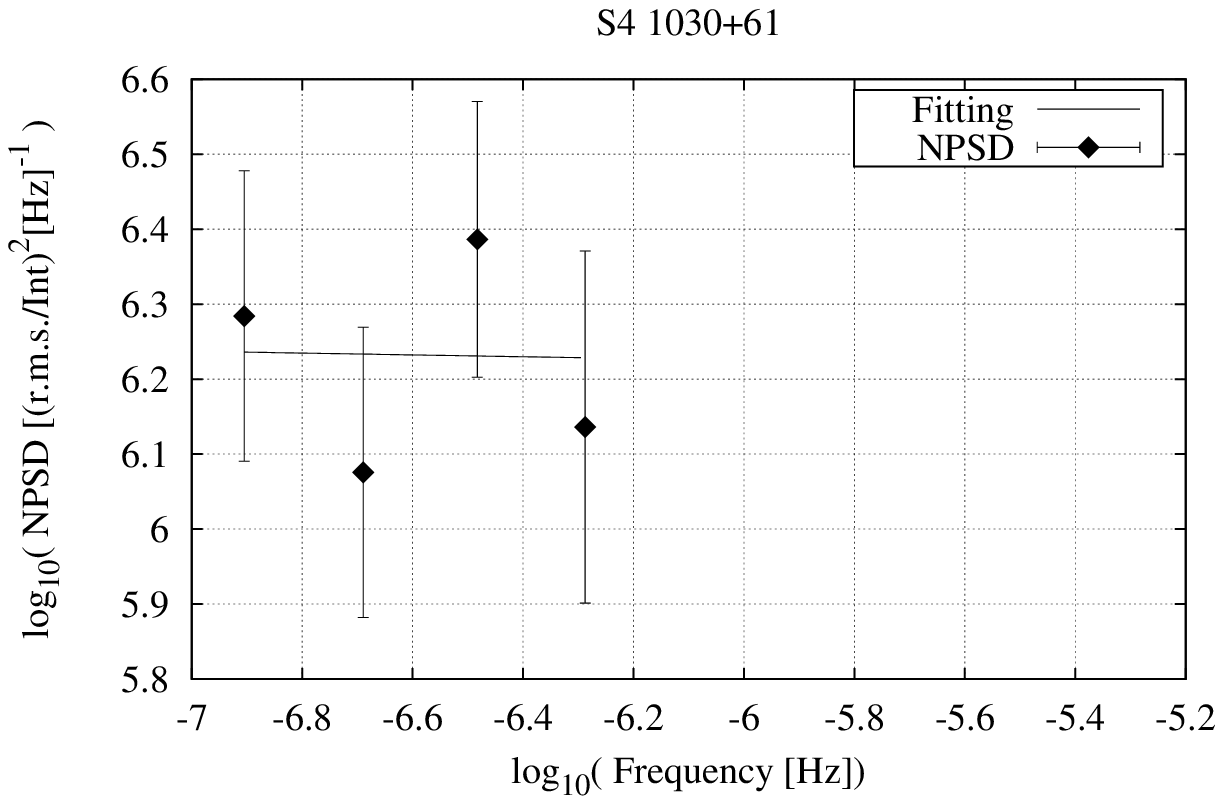}
   \end{center}
  \end{minipage}
 \end{tabular}
 \begin{tabular}{cc}
  \begin{minipage}{0.5\hsize}
   \begin{center}
    \includegraphics[width=6cm]{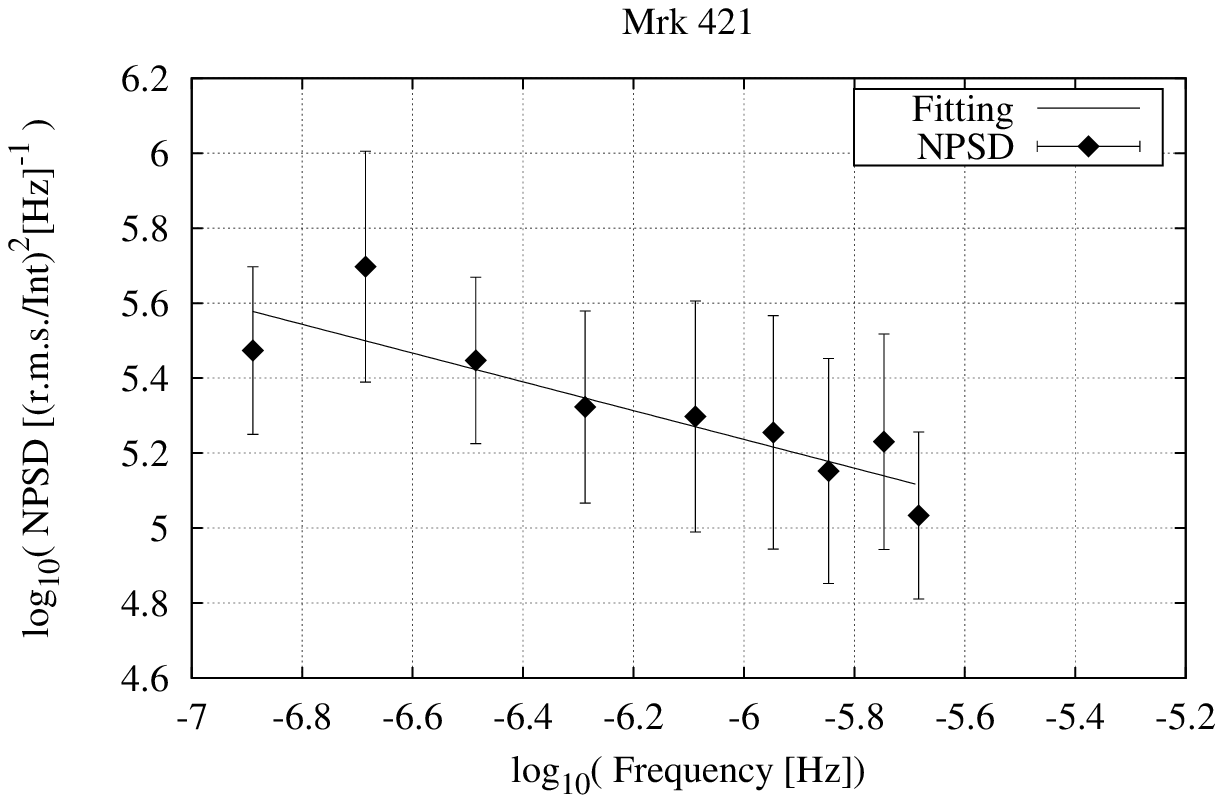}
   \end{center}
  \end{minipage}
  \begin{minipage}{0.5\hsize}
   \begin{center}
    \includegraphics[width=6cm]{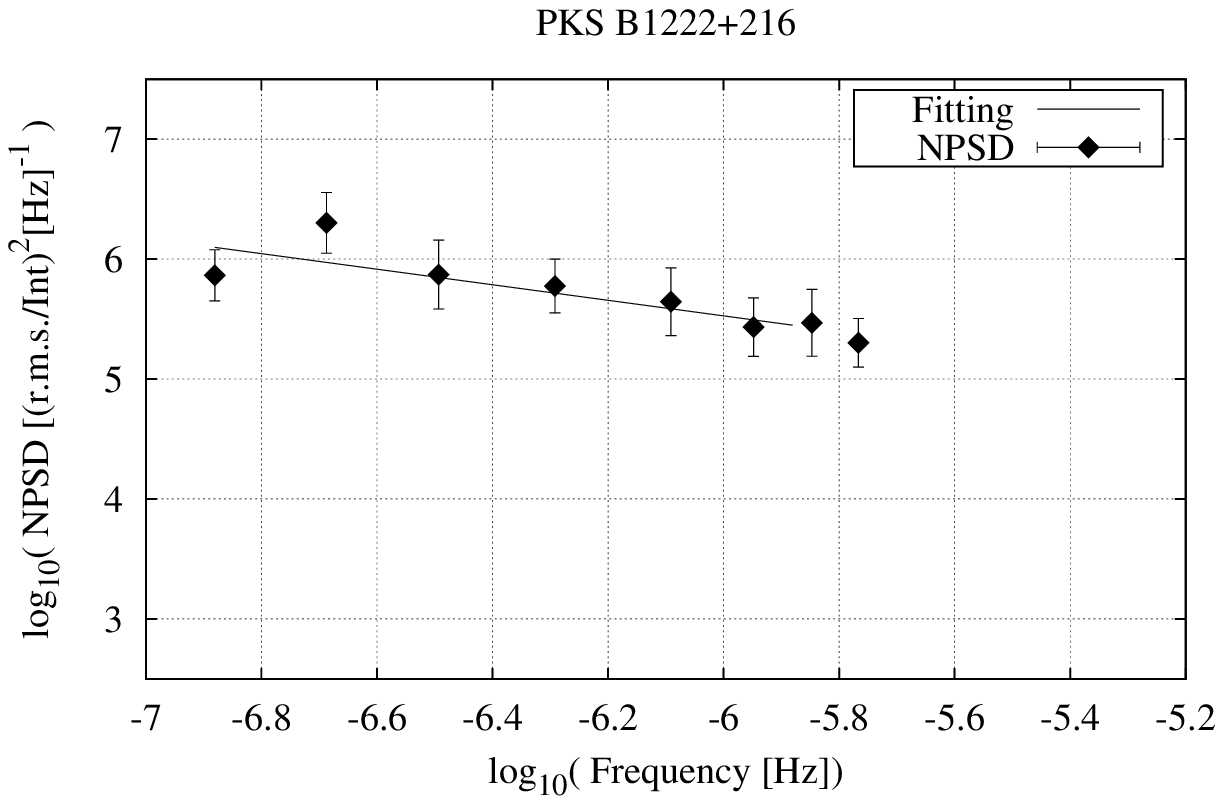}
   \end{center}
  \end{minipage}
 \end{tabular}
 \caption{Plots of NPSD for 15 AGNs.  Lines show the fitting results. Broken lines are 
  drawn when the fitting with a broken power-law gives a better fit (i.e., smaller reduced-$\chi^2$).}
 \label{fig:PSD}
\end{figure}
\clearpage
\setcounter{figure}{0}
\begin{figure}[!htbp]
 \begin{tabular}{cc}
  \begin{minipage}{0.5\hsize}
   \begin{center}
    \includegraphics[width=6cm]{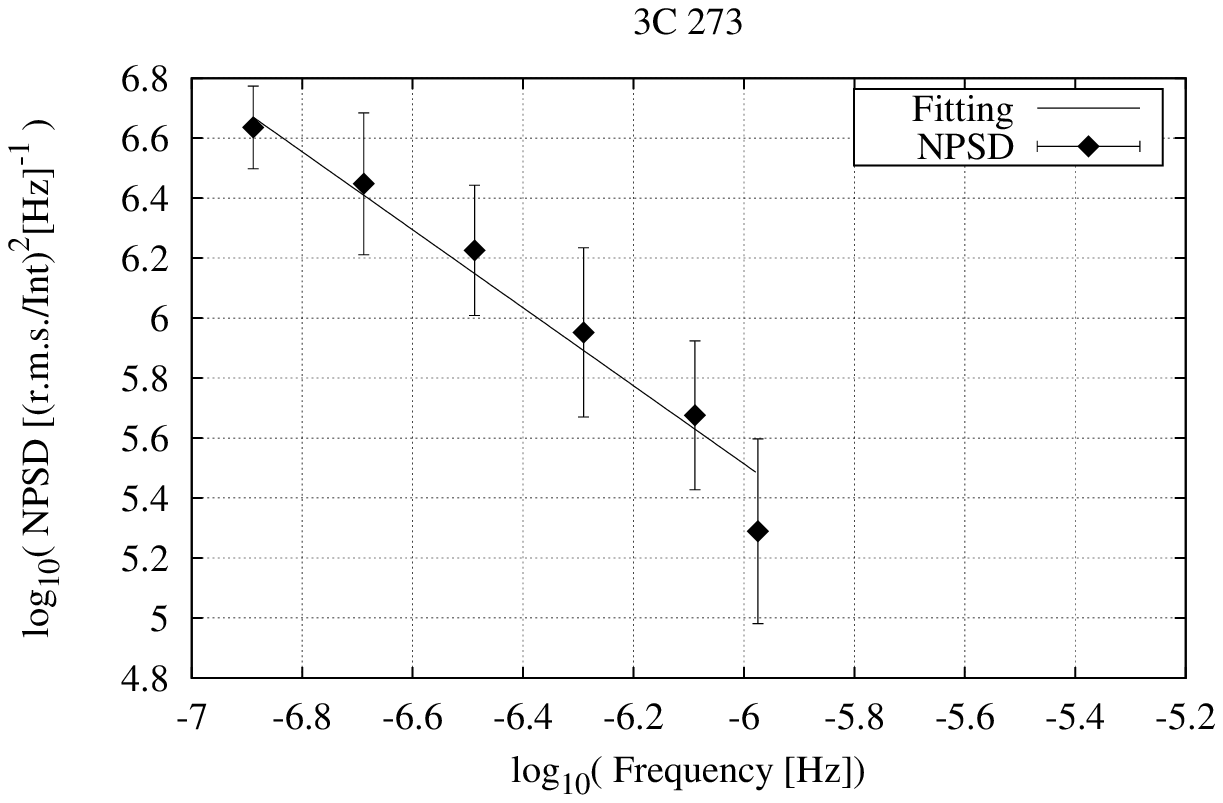}
   \end{center}
  \end{minipage}
  \begin{minipage}{0.5\hsize}
   \begin{center}
    \includegraphics[width=6cm]{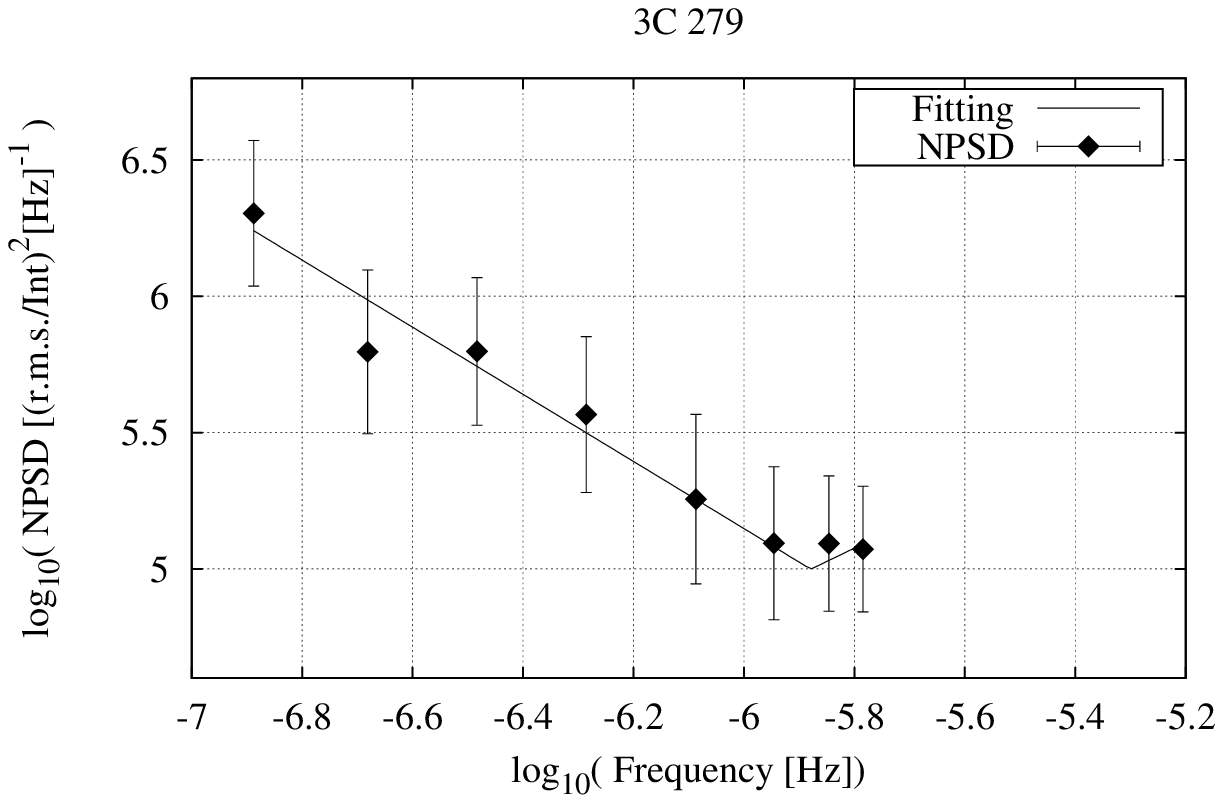}
   \end{center}
  \end{minipage}
 \end{tabular}
 \begin{tabular}{cc}
  \begin{minipage}{0.5\hsize}
   \begin{center}
    \includegraphics[width=6cm]{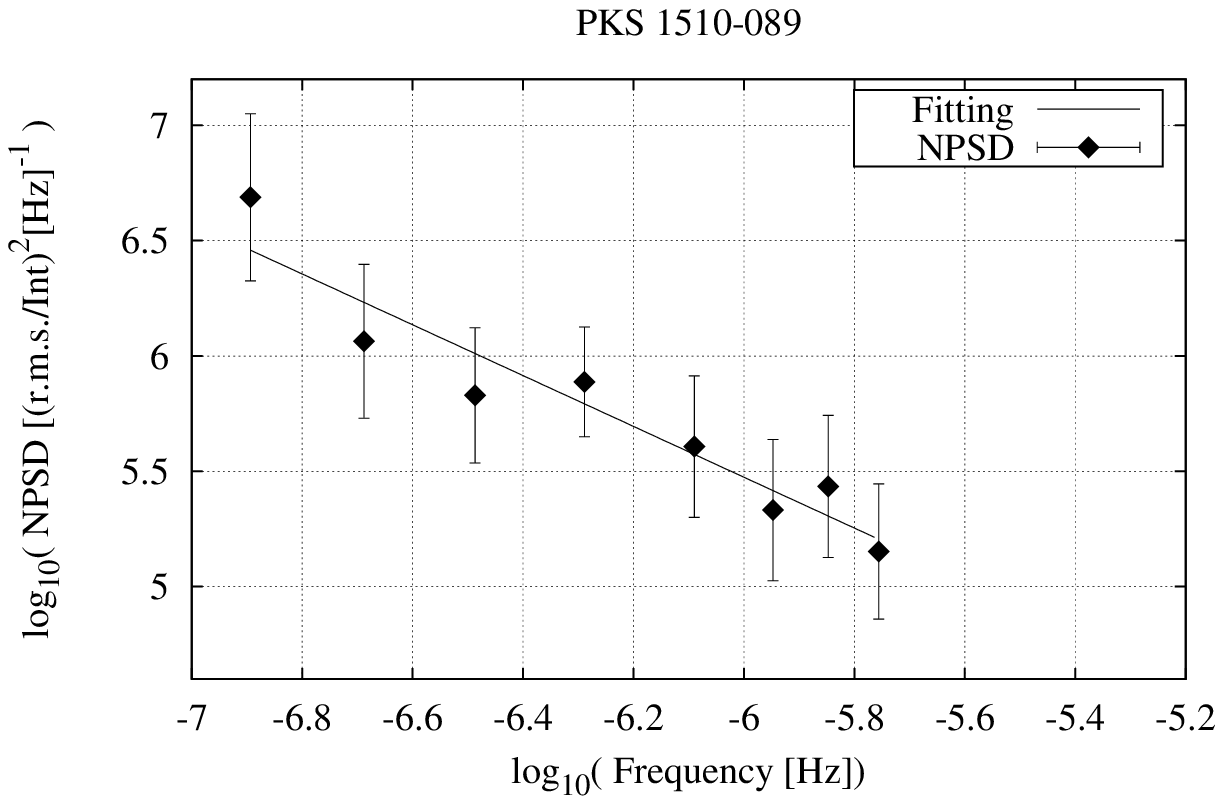}
   \end{center}
  \end{minipage}
  \begin{minipage}{0.5\hsize}
   \begin{center}
    \includegraphics[width=6cm]{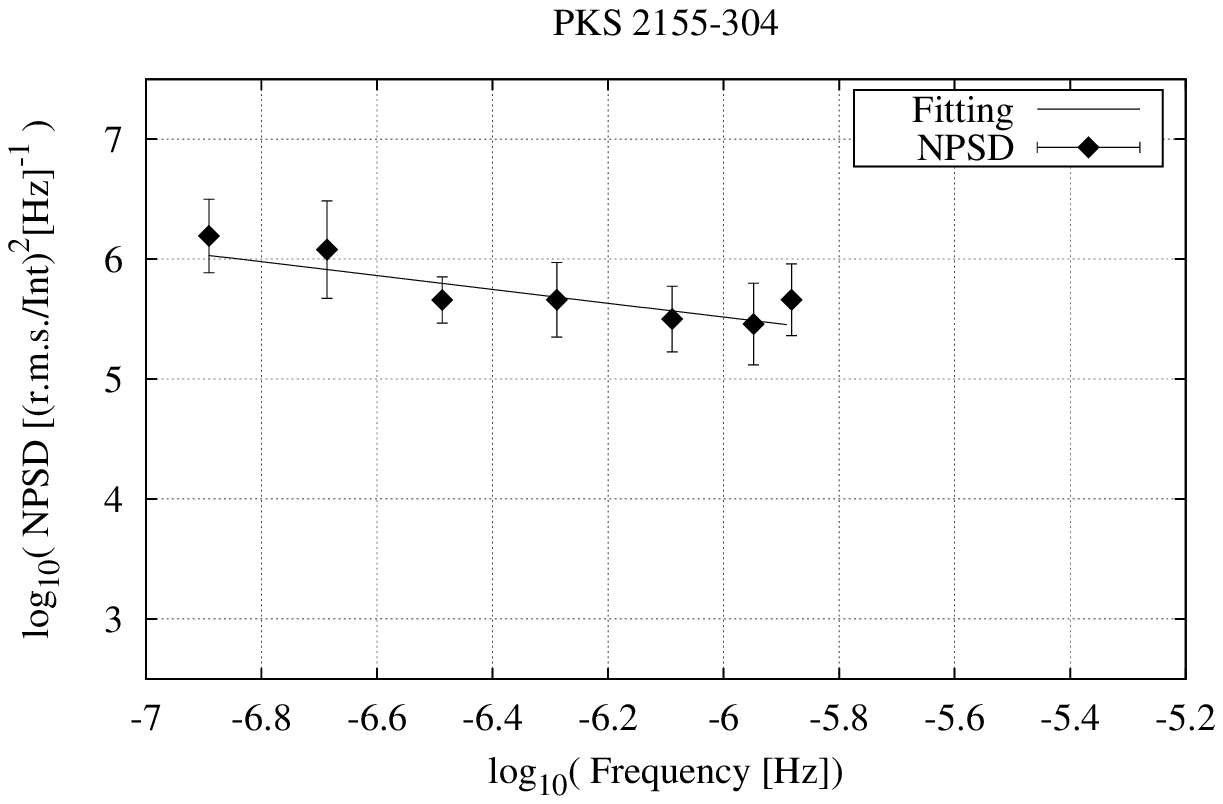}
   \end{center}
  \end{minipage}
 \end{tabular}
 \begin{tabular}{cc}
  \begin{minipage}{0.5\hsize}
   \begin{center}
    \includegraphics[width=6cm]{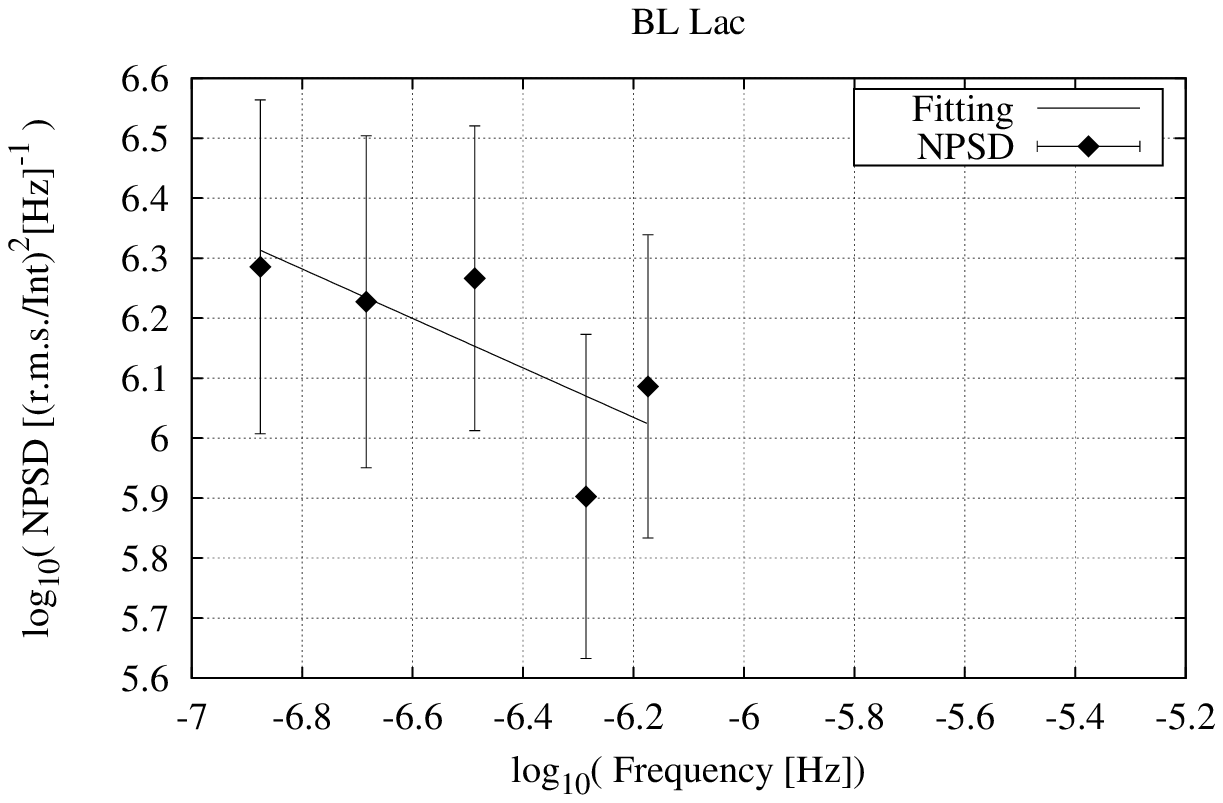}
   \end{center}
  \end{minipage}
  \begin{minipage}{0.5\hsize}
   \begin{center}
    \includegraphics[width=6cm]{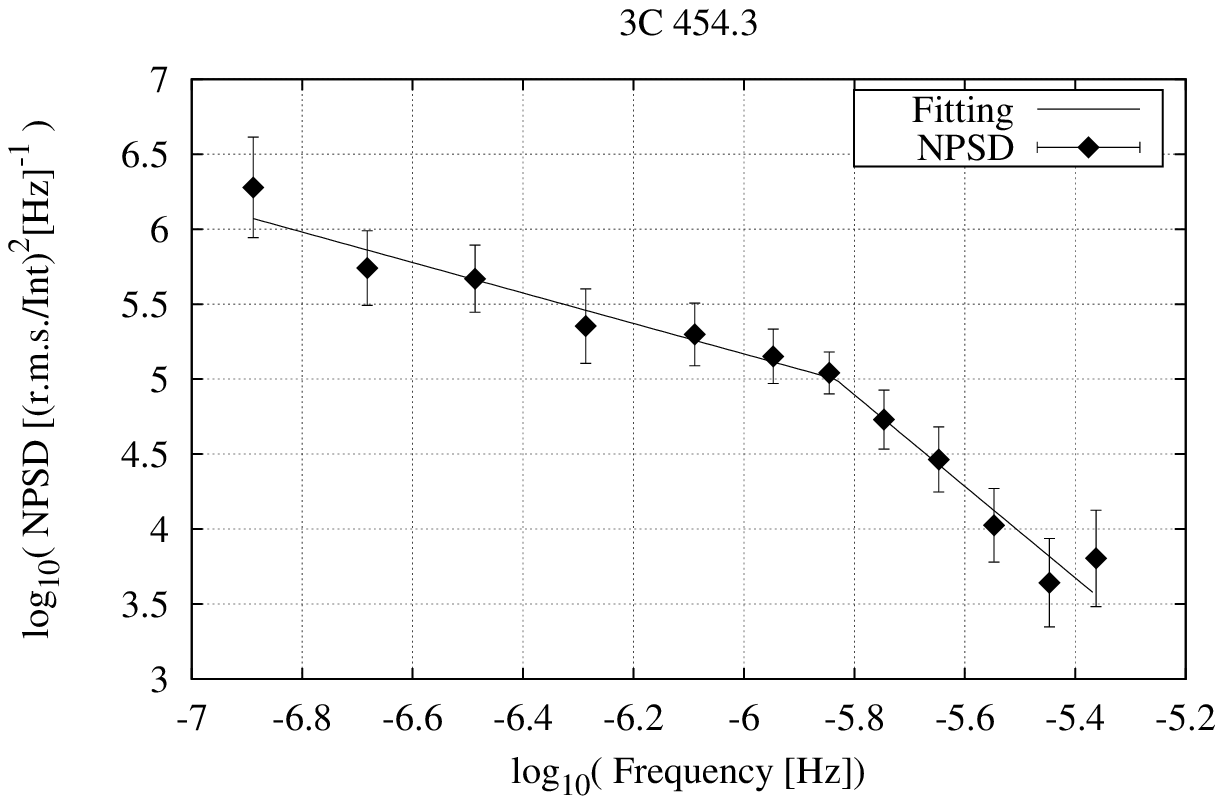}
   \end{center}
  \end{minipage}
 \end{tabular}
 \begin{tabular}{cc}
  \begin{minipage}{0.5\hsize}
   \begin{center}
    \includegraphics[width=6cm]{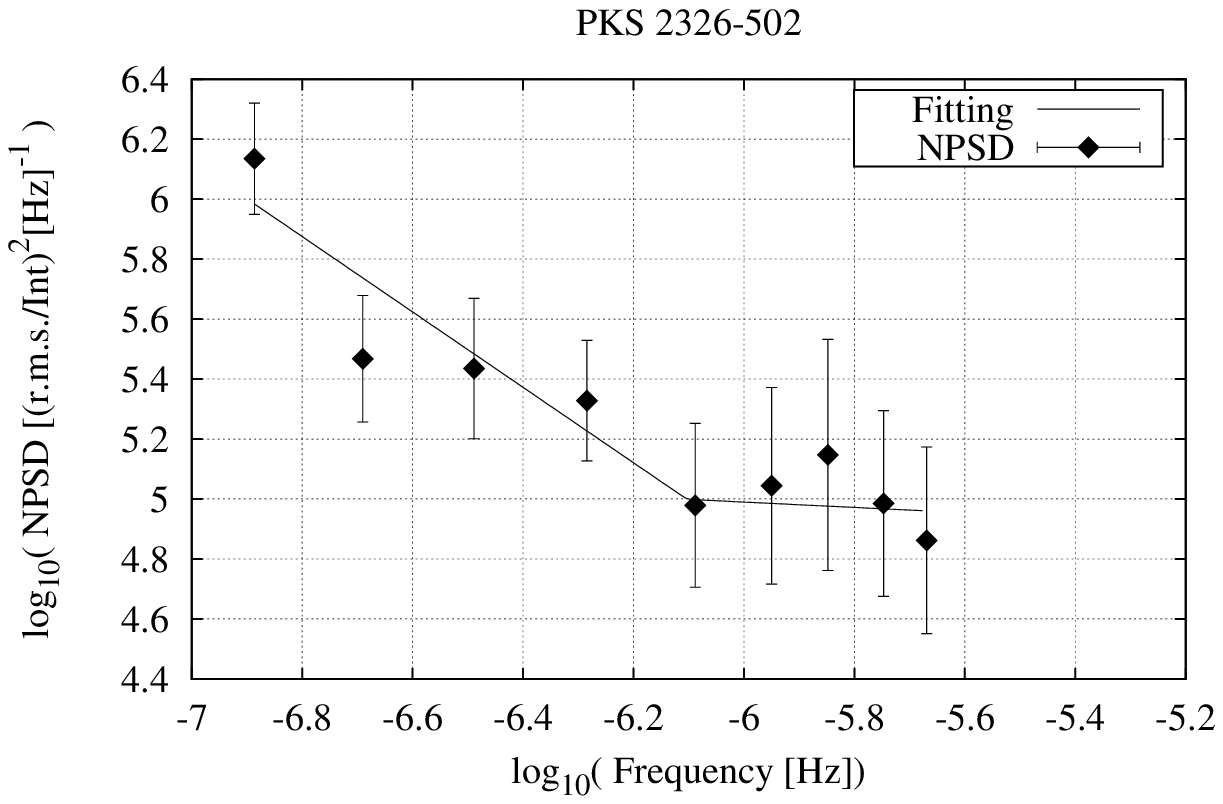}
   \end{center}
  \end{minipage}
 \end{tabular}
\caption{\it Continued.}
\end{figure}

\section{Discussion}

The internal shock model is a popular scenario of blazar
emission as it can explain spectral energy distributions and 
time-lag features \citep[][and references therein]{Boettcher2010}.
\citet{Wehrle2012} studied multiwavelength
variations of the 3C\,454.3 outburst from 2010 November to 2011 January 
with observations by {\sl Herschel}, {\sl Swift}, {\sl Fermi}-LAT,
optical telescopes and submillimeter arrays.
They proposed a model in which turbulent plasma crosses a conical
standing shock in the parsec-scale region of the jet, based on
time-resolved spectral energy distributions for this outburst.
Thus, here we assume the internal shock model
as the emission mechanism for gamma-ray flares
and we interpret the characteristic
time scale that we found in terms of this model.

\citet{Kataoka2001} studied the X-ray variability of three TeV blazars,
Mrk\,421, Mrk\,501 and PKS 2155$-$304, using {\sl ASCA} and {\sl RXTE}
data. In order to interpret the observed characteristic time
scale which they found in their NPSD plots, they assumed a simple model 
based on the internal shock model. They considered two relativistic blobs with 
bulk Lorentz factors $\Gamma$ and $a_0 \Gamma$ ($a_0>1$) ejected at the times 
$t=0$ and $t=\tau_0\, (> 0)$, respectively, and when the second, faster
blob catches up and collide with the first, slower blob, the resulting 
shock generates a high-energy flare.
In this model, the mass of the central black hole, $M_{\rm CBH}$, 
is derived from the variation time-scale, $t_{\rm var}$, as
 \begin{equation}
  M_{\mathrm{CBH}} \simeq 9 \times 10^8 M_{\odot} 
  \frac{t_{\mathrm{var}}}{\mathrm{day}}\frac{10}{k}\frac{a^2_0-1}{2a^2_0} 
  \label{eq:BHmass}
 \end{equation}
where $M_{\odot}$ is the solar mass and $k=c\tau_0/R_g\,\geq 3$ with 
the Schwarzschild radius $R_g$.
They derived $(10^{7}$--$10^{10})M_{\odot}$ 
as the masses of the central black holes of these blazars.

\begin{figure}[htbp]
 \begin{center}
  \includegraphics[width=12cm]{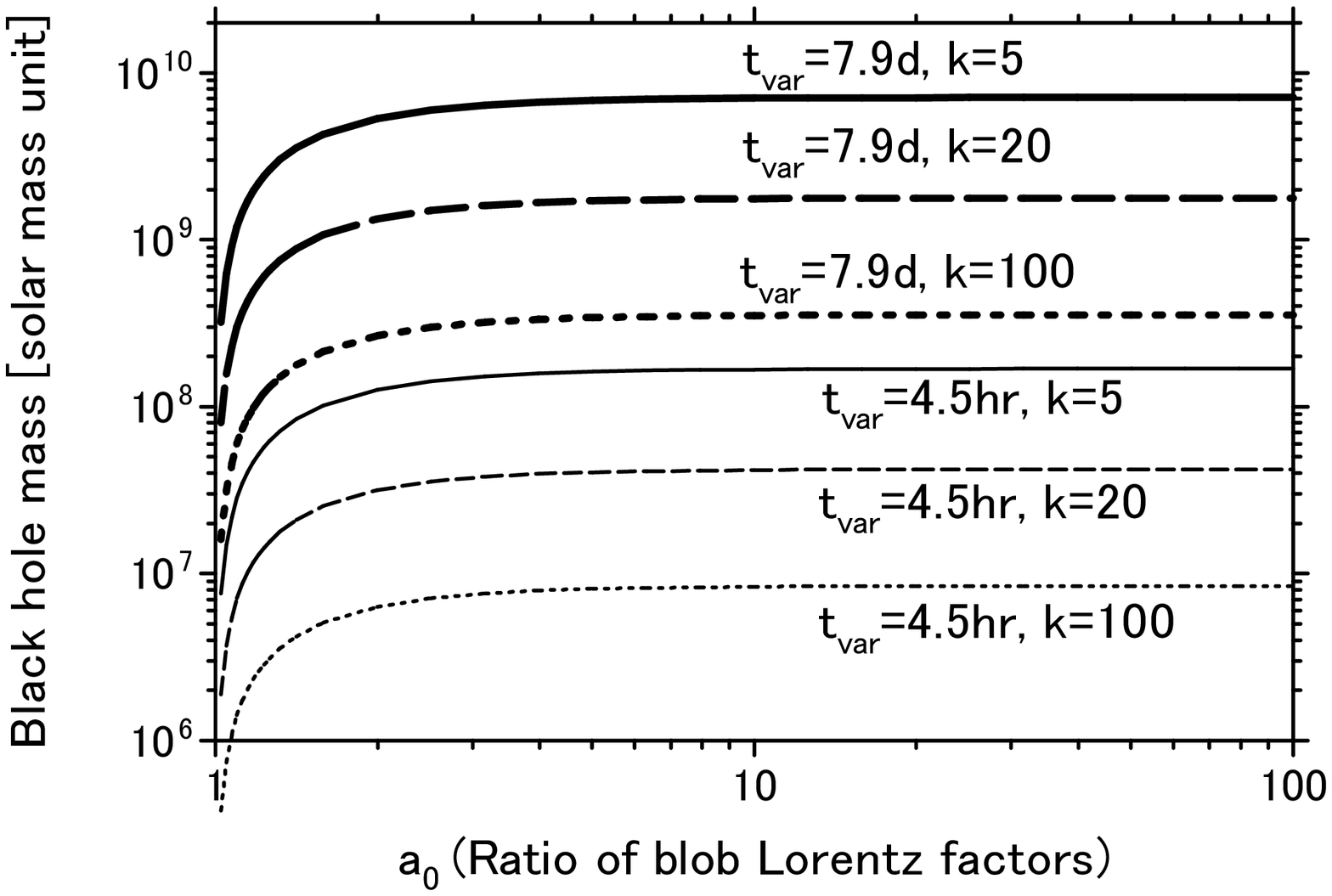}
 \end{center}
 \caption{The central black hole mass as a function of the ratio of blob Lorentz
 factor, $a_0$, assuming the internal shock model for time variation scale 
 $t_{\rm var}$ and some values of 
  $k$ (the light-crossing time in unit of $R_g$ (Schwarzschild radius)$/c$).}
\label{fig:Mass}
\end{figure}

Though this model was devised to explain the X-ray time variability of blazars,
here we assume the same mechanism works for the gamma-ray time
variability%
\footnote{\citet{Jorstad2005} report the ejection of several superluminal knots
based on the VLBA images of 3C\,454.3 sampled during about 3 yr, but these knots
might not be identified as blobs in the model adopted here because
the scale of the phenomena could be different.}%
, and we applied the above equation 
to estimate the central black-hole mass of 3C\,454.3.
We take the variation time-scale as the inverse of the turnover frequency
which we observed in the NPSD plot of 3C\,454.3,
$t_{\rm var}=1/(10^{-5.834} \mathrm{Hz})=6.82 \times 10^{5}$ s $=7.89$ days.
Figure~\ref{fig:Mass} shows the result for several values of model parameters:
$a_0=1$--$100$ and $k=5,\ 20,\ 100$ following \citet{Kataoka2001}.
We can infer the central black hole mass is in the range
$(10^8$--$10^{10}) M_{\odot}$ from this plot in most of the parameter 
space ($a_0\gtrsim 2$, which means the Lorentz factors of colliding blobs
differ significantly).

Alternatively, we can infer the range of the unknown parameter $k$,
or the light crossing time in units of the Schwarzschild radius, by
assuming the central black hole mass estimated by other methods.

3C\,454.3 is one of the most well-known and well-studied gamma-ray sources.
Recently it showed 
two large flares, in 2009 November-December \citep{Ackermann2010, Striani2010}
and 2010 December \citep{Abdo2011}.
\citet{Bonnoli2011} analyzed the 2009 flare and estimated the central
black hole mass by refining the discussion of \citet{Gu2001}, who
used the broad line width and the distance of the broad line region (BLR) 
from the center.
Assuming the broad emission lines 
being produced in clouds which are gravitationally bound and orbiting 
with Keplerian velocities \citep{Dibai1981}, the central black hole mass can be 
given by  $M_{\rm CBH}  R_{\rm BLR}V^2 G^{-1}$, 
where $R_{\rm BLR}$ is the radius of the BLR and $V$ is the velocity of the clouds 
in the BLR. 
\citet{Gu2001} and \citet{Bonnoli2011} estimated 
the central black hole mass of 3C\,454.3 is 
$4 \times 10^9 M_{\odot}$ and $5 \times 10^8 M_{\odot}$, respectively.

\begin{figure}[htbp]
 \begin{center}
  \includegraphics[width=12cm]{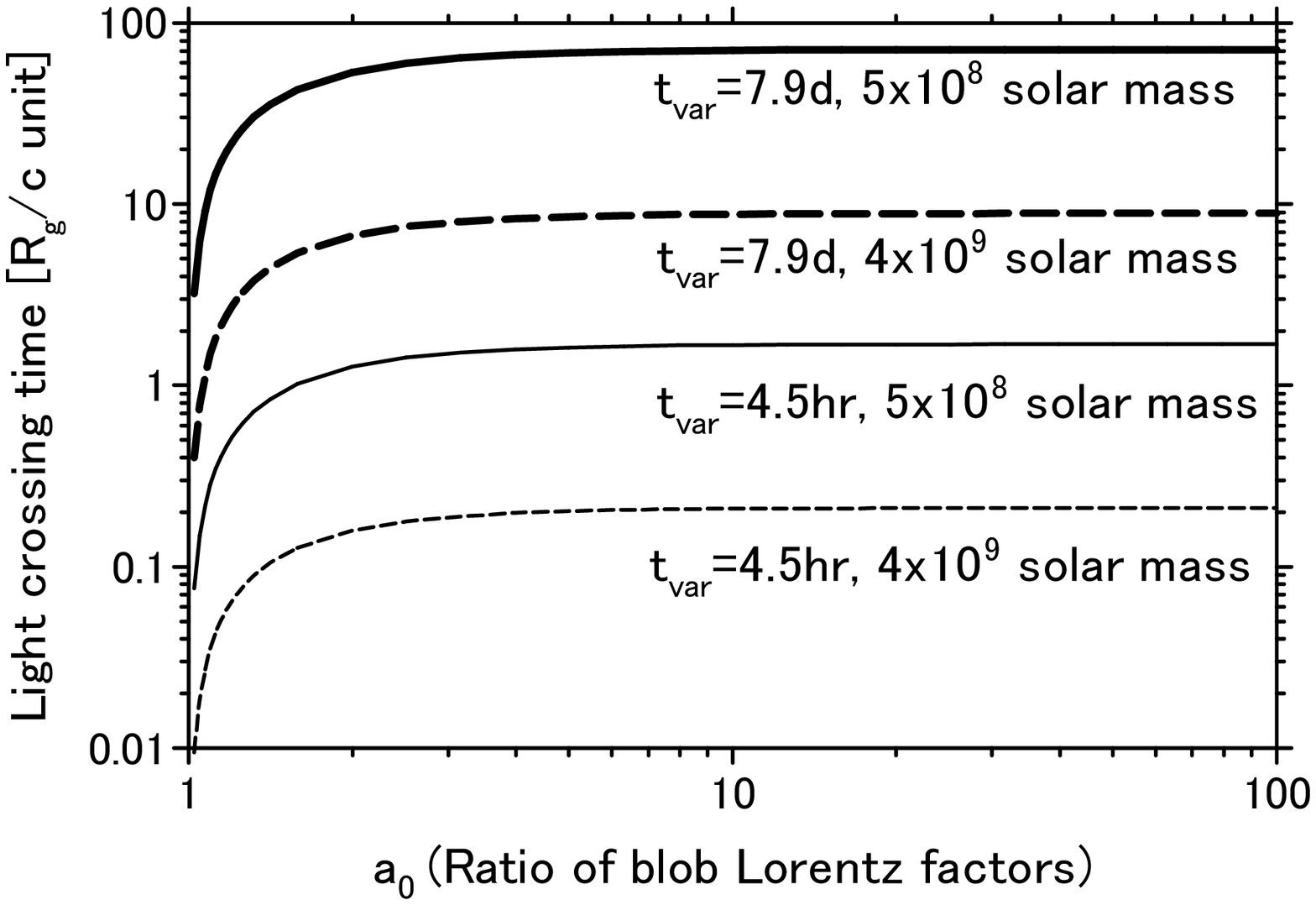}
 \end{center}
 \caption{The light-crossing time $k$ in units of $R_g/c$,
(where $R_g$ is the Schwarzschild radius) 
 as a function of the ratio of blob Lorentz factor, $a_0$, assuming the internal shock model
 for the variation time-scale, $t_{\rm var}$, and central black hole mass. }
\label{fig:Mass_k}
\end{figure}

Figure~\ref{fig:Mass_k} shows the variation of $k=c \tau_0/R_g$ as a function of
the ratio of blob Lorentz factor, $a_0$, for two estimated values of the
central black hole mass.
We see $k$ is in the range of 7 to 70 for $a_0\gtrsim2$.
This parameter range is more restrictive than in the case of general 
consideration for blazar X-ray flares by \citet{Kataoka2001}, where $k$ in the range of
$5\sim 100$ for $t_{\rm var}=1\sim 10$~day and $M_{\rm CBH}=(10^7\sim10^{10})M_\odot$.


The GeV gamma-ray flux during the strong outburst of 3C\,454.3 in 2010 November 
exhibited a very fast variability with the rise time of $4.5\pm 1$\,hr
and the fall time of $14\pm 2$\,hr \citep{Abdo2011}.
We tried to apply the Equation \ref{eq:BHmass} assuming the variation time
scale as this rise time, and the results are shown
in Figures~\ref{fig:Mass} and \ref{fig:Mass_k}.
The central black hole mass inferred from this time scale ($<2\times 10^8M_\odot$,
see Figure~\ref{fig:Mass}) is much smaller than that inferred 
from the characteristic time scale in NPSD, and the
light crossing time ($k<2$, see Figure \ref{fig:Mass_k}) is less than the 
time to cross the last stable orbit of the black hole.
Thus it seems unreasonable to assign the short time scale of
4.5\,hr to the internal shock model under consideration in this paper:
it should be interpreted as, e.g., shocks forming with strong
anisotropic geometries, albeit a low duty cycle (e.g., \cite{Salvati1998}),
or existence of small active regions, inside a larger jet, moving faster
than the rest of the plasma, occasionally pointing toward us \citep{Ghisellini2008}.

\section{Conclusion}

Using the gamma-ray daily light curves
observed by {\it Fermi}-LAT over 3.6 yr, 
we studied the temporal behavior of 15 AGNs by
calculating the NPSDs for each sources.
One source, 3C\,454.3, showed a clear turnover in the NPSD curve
which corresponds to a characteristic time scale of
$6.82 \times 10^{5}$ s.

This time scale can be interpreted as a result of an collision of blobs
in the internal shock of the blazar jet, as discussed by \citet{Kataoka2001}
for the NPSD of X-ray data on blazars.
The time variation scale we found indicates the central black hole mass of 
3C\,454.3 is in the range of $(10^8$--$10^{10})M_{\odot}$.

Alternatively, if we assume the central black hole mass of 3C 454.3 is  
$(0.5$--$4)\times 10^9M_{\odot}$ after \citet{Gu2001} and \citet{Bonnoli2011},
we can infer the time interval emitted the internal shock wave is $7$--$70$ times the
light crossing time of the Schwarzschild radius of the central black hole.

\acknowledgments

We thank the anonymous referee for useful comments
and suggestions.
This work is supported in part by the Grant-in-Aid 
from the Ministry of Education, Culture, Sports, Science
and Technology (MEXT) of Japan, No. 22540315.


\begin{thebibliography}{}

\bibitem[Abdo et al.(2010a)]{Abdo2010a}
Abdo, A.A. et al., 2010a, \nat, 463, 919

\bibitem[Abdo et al.(2010b)]{Abdo2010b}
Abdo, A.A. et al., 2010b, \apj, 722, 520


\bibitem[Abdo et al.(2011)]{Abdo2011}
Abdo, A.A. et al., 2011, \apjl, 733, L26

\bibitem[Ackermann et al.(2011)]{Ackermann2010}
Ackermann, M. et al., 2010, \apjl, 721, 1383

\bibitem[Aharonian et al.(2007)]{Aharonian2007}
Aharonian. F., et al., 2007, \apj,  664, L71

\bibitem[Aleksi\'{c} et al.(2011)]{Aleksic2011}
Aleksi\'{c}, B., et al., 2011, \apj, \apj, 730, L8

\bibitem[Atwood et al.(2009)]{Atwood2009}
Atwood, W.B., Abdo, A.A. et al., 2009, \apj, 697, 1071

\bibitem[Bonnoli et al.(2011)]{Bonnoli2011}
Bonnoli, G. et al., 2011, \mnras, 410, 368

\bibitem[Barr \& Mushotzky(1986)]{Barr1986}
Barr, P. and Mushstzky, R.F., \nat, 1986, 320, 421

\bibitem[B\"{o}ttcher \& Dermer(2010)]{Boettcher2010}
B\"{o}ttcher, M. and Dermer, C.D., 2010, \apj, 445 

\bibitem[Chatterjee et al.(2012)]{Chatterjee2012}
Chatterjee, R. et al., 2012, \apj, 749, 191

\bibitem[Czerny et al.(2001)]{Czerny2001}
Czerny, B. et al., 2001, \mnras, 325, 865

\bibitem[Dibai(1981)]{Dibai1981}
Dibai, E.A., 1981, \sovast, 24, 389

\bibitem[Fossati et al.(1998)]{Fossati1998}
Fossati, G., Maraschi, L., Celotti, A., Comastri, A., Ghisellini, G.
1998, \mnras, 299, 433

\bibitem[Ghisellini \& Tavecchio(2008)]{Ghisellini2008}
Ghisellini, G. \& Tavecchio, F., 2001, \mnras, 386, L28

\bibitem[Gu et al.(2001)]{Gu2001}
Gu, M. et al., 2001, \mnras, 327, 1111

\bibitem[Hayashida et al.(1998)]{Hayashida1998}
Hayashida, K., Miyamoto, S. et al., 1998, \apj, 500, 642

\bibitem[Jorstad et al.(2005)]{Jorstad2005}
Jorstad, S.G. et al., 2005, \aj, 130, 1418

\bibitem[Kataoka et al.(2001)]{Kataoka2001}
Kataoka, J. et al., 2001, \apj, 560, 659

\bibitem[Lawrence et al.(1987)]{Lawrence1987}
Lawrence, A. et al., 1987, \nat, 325, 694

\bibitem[McHardy \& Czerny(1987)]{McHardy1987}
McHardy, I. and Czerny, B., 1987, \nat, 325, 696


\bibitem[Miyamoto et al.(1994)]{Miyamoto1994}
Miyamoto, S. et al., 1994, \apj, 435, 398

\bibitem[Nolan et al.(2012)]{Nolan2012}
Nolan, P.L., Abdo, A.A. et al., 2012, \apjs, 199, 31

\bibitem[Nowak et al.(2012)]{Nowak2012}
Nowak, N. et al., in AIP Conference Proceedings Vol. 1505
(eds. Aharonian, F.A., Hofmann, W. \& Riger, F.M.), 2012, 518

\bibitem[Salvati, Spada and Pacini(1998)]{Salvati1998}
Salvati, M., Spada, M., \& Pacini, F. 1998, \apj, 495, L19

\bibitem[Striani et al.(2010)]{Striani2010}
Striani, E. et al., 2010, \apj, 718, 455

\bibitem[Ulrich, Maraschi \& Urry(1997)]{Ulrich1997}
Ulrich, M.-H., Maraschi, L. and Urry, C. M., 1997, \araa, 35, 445

\bibitem[Urry \& Padovani(1995)]{Urry1995}
Urry, C. M., and Padovani, P. 1995, \pasp, 107, 803

\bibitem[Vaughan, Fabian and Nandra (2003)]{Vaughan2003}
Vaughan, S. Fabian, A.C. and Nandra, K., 2003, \mnras, 339, 1237

\bibitem[Wehrle et al.(2012)]{Wehrle2012}
Wehrle, A.E. et al., 2012 \apj,, 758, 72

\end{thebibliography}
\end{document}